\documentclass[a4 paper,12pt]{article}
\usepackage{amsmath}
\usepackage{amsfonts}
\usepackage{amssymb}
\usepackage{graphicx}
\usepackage{array}
\usepackage{multirow}
\usepackage{setspace}
\usepackage{algorithm}
\usepackage[noend]{algpseudocode}
\usepackage[margin=1 in]{geometry}
\usepackage{float}
\usepackage[utf8]{inputenc}
\usepackage{graphicx}
\usepackage{tikz}% Required for inserting images
\usetikzlibrary{decorations.pathmorphing}
\usepackage{xcolor}
\newcolumntype{P}[1]{>{\centering\arraybackslash}p{#1}}
\newcolumntype{M}[1]{>{\centering\arraybackslash}m{#1}}
\usepackage{cite}
\usepackage{hyperref}
\usepackage{authblk}
\renewcommand\thefootnote{}

\begin{document}
\begin{center}
{\bf\large{Faddeev-Jackiw Approach to Classical Constrained Systems} }

\vskip 1.5 cm

{\sf{ \bf Shaza Abdul Majid, Ansha S Nair and Saurabh Gupta}}\\
\vskip .1cm
{\it Department of Physics, National Institute of Technology Calicut,\\ Kozhikode - 673 601, Kerala, India}\\
\vskip .15cm
{E-mail: {\tt anshsuk8@gmail.com}}
\end{center}
\vskip 1cm

\noindent
\textbf{Abstract:} We accomplish the quantization of a few classical constrained systems \textit{\'{a} la} (modified) Faddeev-Jackiw formalism. We analyze the constraint structure and obtain basic brackets
of the theory. In addition, we disclose the gauge symmetries within the symplectic framework.
We also provide an interpretation for Lagrange multipliers and outline a MATLAB
implementation algorithm for symplectic formulation.

\vskip 1.5cm
\noindent
%\textbf{PACS Nos:} 11.15.-q, 11.10.Ef, 11.30.-j
\vskip 1cm
\noindent
%\textbf{Keywords:} Modified Faddeev-Jackiw formalism; FLPR model; Constrained systems; Gauge symmetries.
\clearpage

\section{Introduction}
\medskip

In classical mechanics, constraints play a significant role in the dynamics
of a system. These constraints can be imposed externally on the system,
or can be inherent, originating from the system itself due to the singularity
of its Lagrangian. These singular systems can not be quantized using
conventional methods. To address this issue, Dirac introduced a method in which quantization is accomplished by categorizing constraints into different classes and constructing Dirac brackets\cite{dirac2001lectures}.
While the Dirac formalism serves its purpose, it contains several intricate logical steps, and can be tedious in some cases\cite{universe8030171}. This complexity has given rise to alternative approaches for analyzing constrained systems. 

The Faddeev-Jackiw formalism is one such alternative approach. This approach utilizes the symplectic structure of phase space and  streamlines the treatment of constraints, making it relatively easier to handle systems with multiple constraints \cite{PhysRevLett.60.1692, Barcelos-Neto:1991dhe, S:2021msx}. The core idea of this symplectic quantization is to rewrite the Lagrangian in first-order form and construct the symplectic two-form from its kinetic part. Constraints are obtained with the aid of zero-modes of the symplectic two-forms, and consistency conditions are applied to deduce new constraints. This is an iterative process and it is continued until no new constraints are generated\cite{Foussats1997}. These constraints are incorporated into the Lagrangian and the symplectic two-form matrix is modified 
till it becomes invertible. This non-singular symplectic matrix gives the basic
brackets of the theory \cite{PhysRevLett.60.1692}.
The broad applicability of the Faddeev-Jackiw method has led to various studies exploring its use in the quantization of superconducting systems, non-abelian systems, four-dimensional BF theory, particle on a torus knot, anti self-dual Yang-Mills theory, classical mechanical systems, and in the path integral framework (see, e.g. \cite{HUANG20102140, doi:10.1142/S0217732393003810, Gupta:2016qwv,  ESCALANTE2016375,  Anjali:2019ocq, S:2021zob, doi:10.1142/S0217732323501869, Paulin-Fuentes:2023zya, Toms:2015lza } for details).

This work intends to apply the Faddeev-Jackiw formalism to analyze classical dynamical systems with singular Lagrangians. All systems discussed here consist of elements commonly found in classical mechanics problems, such as masses, springs and rods. These systems have previously been studied in the context of the Dirac-Bergmann method \cite{brown2023singular} and Hamilton-Jacobi method \cite{RomeroHernandez2025}. The main motive of this work is to obtain the basic brackets of given systems using a geometrically motivated approach - the Faddeev-Jackiw method. Further, we wish to examine the equivalence of the two methods in the context of classical constrained systems (i.e., Dirac and Faddeev-Jackiw) by comparing our findings with \cite{brown2023singular}.

This paper takes the following form. Section \ref{2} introduces the Faddeev-Jackiw formalism in a nutshell. The application of this formalism to various classical constrained systems is carried out in section \ref{msr}. Subsection \ref{sec:1} deals with a system of four masses fixed at the midpoints of four massless, freely extensible rods. In \ref{sub2} we consider three identical masses that slide without friction on a ring. A system made up of three pairs of massless pulleys is analyzed in section \ref{sub3}. Finally, we summarize our results in section \ref{con}. 

In Appendix A, we explain the ``physical" interpretation of the Lagrange multipliers. We also sketch the outline of MATLAB implementation of symplectic formulation in Appendix B.

\section{The Faddeev-Jackiw Formalism: A Brief Review}
\label{2}
Faddeev-Jackiw formalism is a geometrically motivated approach for the quantization of constrained systems. This formalism constructed upon a first-order Lagrangian and utilizes the symplectic structures of phase space. We begin with a first-order Lagrangian given as\cite{PhysRevLett.60.1692}
\begin{equation}
    L^{(0)}=a_{i}^{(0)}\dot{\zeta}^{i(0)}-V^{(0)}(\zeta),
\end{equation}
where $i=1,2,...n.$ Here, $\zeta_{i}^{(0)}$ denotes the generalized coordinates in an $n$-dimensional configuration space, $a_{i}^{(0)}$ and $V^{(0)}(\zeta)$ represents the canonical one-form and symplectic potential, respectively. The equations of motion can be derived in the following manner
\begin{equation}
    f_{ij}^{(0)} \dot{\zeta}^j = \frac{\partial V^{(0)}(\zeta)}{\partial \zeta^i},
    \label{eq:1}
\end{equation}
where $f_{ij}^{(0)}$ is the symplectic two-form matrix and can be constructed by
\begin{equation}
     f_{ij}^{(0)} = \frac{\partial a_j^{(0)}(\zeta)}{\partial \zeta_i^{(0)}} - \frac{\partial a_i^{(0)}(\zeta)}{\partial \zeta_j^{(0)}}.
\end{equation}
If this symplectic matrix $(f_{ij}^{(0)})$ is singular, it implies that the underlying theory is a constrained theory\cite{Anjali:2019ocq}. In this case, all constraints are obtained from the zero-modes of the singular symplectic matrix $(f_{ij}^{(0)})$ in an iterative manner. Let $\nu_{\alpha}^{(0)}$ are the zero-modes of the symplectic matrix $f_{ij}^{(0)}$, then the constraints $(\Omega^{(0)}_{\alpha})$ can be procured as
\begin{equation}
     \Omega^{(0)}_{\alpha} = \left( \nu^{(0)}_{\alpha} \right)^T \frac{\partial V^{(0)}(\zeta)}{\partial \zeta^{(0)}}=0,
\end{equation}
where $\alpha=1,2,...,m.$ Using the modified Faddeev-Jackiw method, the new constraints are deduced by means of consistency condition on constraints analogous to Dirac-Bergmann algorithm as follows\cite{HUANG2008438}:
\begin{equation}
    \dot{\Omega}^{(0)}_{\alpha} = \frac{\partial \Omega^{(0)}_{\alpha}}{\partial \zeta^i} \dot{\zeta}^i = 0.
\end{equation}
Combining the above equation with (\ref{eq:1}), we obtain
\begin{equation}
        f_{kj}^{(1)} \dot{\zeta}^j = Z_k(\zeta),
        \label{eq:2}
\end{equation}
where,
\begin{equation}
    f_{kj}^{(1)}=
    \begin{bmatrix}
        f_{ij}^{(0)}\\
        \frac{\partial\Omega^{(0)}}{\partial\zeta^i}
    \end{bmatrix},\,\,Z_k(\zeta)=
    \begin{bmatrix}
        \frac{\partial V^{(0)}(\zeta)}{\partial\zeta^i}\\
        0
    \end{bmatrix}.
    \label{eq:3}
\end{equation}
Multiplying the zero-modes of $f_{kj}^{(1)}$ to both sides of (\ref{eq:2}) under the condition $\Omega^{(0)}_\alpha=0$, results in two possibilities. If the resulting relation gives identity, then there are no further constraints in the theory\cite{doi:10.1142/S0217732323501869}. If not, it gives rise to new constraints.

Now we incorporate the obtained constraints into the Lagrangian using Lagrange multipliers. We identify the new set of symplectic variables of the theory and construct the first-iterated symplectic matrix ($f_{ij}^{(1)}$). If the matrix is still singular and there are no further constraints in the theory, it indicates the presence of a gauge symmetry. The zero-modes of the respective singular matrix will act as the generators of gauge transformations\cite{doi:10.1142/S0217732393003810}. In this case, in order to quantize the theory, we introduce a gauge fixing condition into the Lagrangian and construct the two-form symplectic matrix. The components of the inverse of this matrix directly gives the basic brackets of the theory. The form of gauge transformations can be written as
\begin{equation}
    \delta \zeta_k^{(i)} = \nu_k^{(i)} \kappa,
\end{equation}
where $i$ represents the level of iteration where the gauge symmetries disclosed and $\kappa(t)$ denotes the time-dependent infinitesimal gauge parameter.

\section{Quantization of Classical Systems}
\label{msr}
\subsection{Masses, Rods and Springs}
\label{sec:1}
\vskip .1cm
\noindent
In this section we analyze a system of four masses with springs attached to them. The masses are fixed at the midpoints of four massless, freely extensible rods and connected to the ceiling with the help of springs (cf. \cite{brown2023singular} for detail description). 
The Lagrangian of the system is given by
\begin{equation}\label{L}
L=\frac{1}{8}m[(\Dot{y_1}+\Dot{y_2})^2+(\Dot{y_2}+\Dot{y_3})^2+(\Dot{y_3}+\Dot{y_4})^2+(\Dot{y_1}+\Dot{y_4})^2]-V(y),
\end{equation}
where
\begin{equation}
\begin{aligned}
V(y) &= mg(y_1+y_2+y_3+y_4)+\frac{k}{8}[(2a-y_1-y_2)^2\\
&+(2a-y_2-y_3)^2+(2a-y_3-y_4)^2+(2a-y_4-y_1)^2].
\end{aligned}
\end{equation}
Here $y_1, y_2, y_3$ and $y_4$ are the coordinates of the connection points where the springs are attached to the rods, $k$ represents the spring constant and $a=h-l$, where $h$ is the height of the ceiling and $l$ denotes the relaxed length of each spring \cite{brown2023singular}.
The canonical momenta are defined as
\begin{equation}
    \begin{aligned}
        &p_1=\frac{m}{4}(\Dot{y_4}+2\Dot{y_1}+\Dot{y_2}), \quad
        p_2=\frac{m}{4}(\Dot{y_1}+2\Dot{y_2}+\Dot{y_3}),\\
        &p_3=\frac{m}{4}(\Dot{y_2}+2\Dot{y_3}+\Dot{y_4}), \quad
        p_4=\frac{m}{4}(\Dot{y_3}+2\Dot{y_4}+\Dot{y_1}),
    \end{aligned}
\end{equation}
which leads to the following constraint in the theory
\begin{equation}
    \phi \equiv p_1-p_2+p_3-p_4=0.
\end{equation}
The canonical Hamiltonian $(H_{c})$ of the system can be obtained as\cite{brown2023singular}
\begin{equation}
    H_c = \frac{1}{2m}\Big[\frac{5}{2}(p_1^2+p_2^2+p_3^2+p_4^2)-2(p_1p_2+p_2p_3+p_3p_4+p_4p_1)+p_1p_3+p_2p_4\Big]+ V(y).
\end{equation}
To study the system by means of Faddeev-Jackiw formalism, we express the above Lagrangian \eqref{L} in its first-order form, as follows:
\begin{equation}
    L^{(0)} = p_1\Dot{y_1}+p_2\dot{y_2}+p_3\dot{y_3}+p_4\dot{y_4}-V^{(0)},
\end{equation}
where $V^{(0)}=H_{c}$.
The constraint $\phi$ is introduced into the Lagrangian using a Lagrange multiplier $\lambda$ as
\begin{equation}
L^{(1)} = p_1\Dot{y_1}+p_2\dot{y_2}+p_3\dot{y_3}+p_4\dot{y_4}+\dot{\lambda}(p_1-p_2+p_3-p_4)-V^{(1)},
\end{equation}
where $V^{(1)}=V^{(0)}\vert_{\phi=0}$. To obtain the constraint structure of the system, we construct the first-iterated symplectic matrix $f_{ij}^{(1)}$. The set of first-iterated symplectic variables of the theory are identified as
\begin{equation}
    \zeta^{(1)}=\{y_1, p_1, y_2, p_2, y_3, p_3, y_4 ,p_4, \lambda\},
\end{equation}
and corresponding components of the symplectic one-forms are given below:
\begin{equation}
\begin{aligned}
&a_{y_{1}}^{(1)}=p_1, \quad a_{p_{1}}^{(1)}=0, \quad a_{y_{2}}^{(1)}=p_2, \quad a_{p_{2}}^{(1)}=0,\quad 
a_{y_{3}}^{(1)}=p_3,\\ \quad &a_{p_{3}}^{(1)}=0, \quad a_{y_{4}}^{(1)}=p_4, \quad a_{p_{4}}^{(1)}=0,\quad 
a_\lambda^{(1)}=p_1-p_2+p_3-p_4.
\end{aligned}
\end{equation}
The symplectic matrix $f_{ij}^{(1)}$ takes the following form
\begin{equation}
f_{ij}^{(1)}=
\begin{bmatrix}
    0 & -1 & 0 & 0 & 0& 0& 0&0&0\\
    1&0&0&0&0&0&0&0&1\\
    0&0&0&-1&0&0&0&0&0\\
    0&0&1&0&0&0&0&0&-1\\
    0&0&0&0&0&-1&0&0&0\\
    0&0&0&0&1&0&0&0&1\\
    0&0&0&0&0&0&0&-1&0\\
    0&0&0&0&0&0&1&0&-1\\
    0&-1&0&1&0&-1&0&1&0
\end{bmatrix}.
\end{equation}
This matrix is singular which indicates the presence of constraints in the theory. %The symplectic equations of motion in terms of the symplectic matrix $(f_{ij}^{(1)})$ can be derived as follows
%\begin{equation}
   %f_{ij}^{(1)}\dot{\zeta}^{j}=\frac{\partial V^{(1)}(\zeta)}{\partial\zeta^{(i)}}
%\end{equation}
The zero-mode  $(\nu^{(1)})$ of the matrix turns out to be
\begin{equation}
    (\nu^{(1)})^T=(-\alpha, 0, \alpha, 0, -\alpha, 0, \alpha, 0, \alpha).
\end{equation}
Contracting this zero-mode with the gradient of first-iterated symplectic matrix $(\frac{\partial V^{(1)}(\zeta)}{\partial\zeta^{(i)}})$ yields identity, confirming that no new constraints arise. The singularity of the matrix with no additional constraints indicates the presence of a gauge symmetry. The gauge condition we choose to quantize the theory is $y_2-y_1=0$, requiring the rod between $y_1$ and $y_2$ to be horizontal \cite{brown2023singular}.
We now incorporate this into the Lagrangian using the Lagrange multiplier ($\rho$) as follows
\begin{equation}
    L^{(2)}=p_1\dot{y_1}+p_2\dot{y_2}+p_3\dot{y_3}+p_4\dot{y_4}+\dot{\lambda}(p_1-p_2+p_3-p_4)+\dot{\rho}(y_2-y_1)-V^{(2)},
\end{equation}
where $V^{(2)}=V^{(1)}\vert_{(y_2-y_1)=0}$.
The set of second-iterated symplectic variables are 
\begin{equation}
    \zeta^{(2)}=\{y_1,p_1,y_2,p_2,y_3,p_3,y_4,p_4,\lambda,\rho\},
\end{equation}
and the respective symplectic one-forms are 
\begin{equation}
\begin{aligned}
&a_{y_{1}}^{(2)}=p_1, \quad a_{p_{1}}^{(2)}=0, \quad a_{y_{2}}^{(2)}=p_2, \quad a_{p_{2}}^{(2)}=0,\quad 
a_{y_{3}}^{(2)}=p_3, \quad a_{p_{3}}^{(2)}=0,\\&a_{y_{4}}^{(2)}=p_4, \quad a_{p_{4}}^{(2)}=0,
\quad a_\lambda^{(2)}=p_1-p_2+p_3-p_4, \quad  a_\rho^{(2)}=y_2-y_1.
\end{aligned}
\end{equation}
The second-iterated symplectic matrix can be calculated as follows
\begin{equation}
f_{ij}^{(2)}=
    \begin{bmatrix}
    0 & -1 & 0 & 0 & 0& 0& 0&0&0&-1\\
    1&0&0&0&0&0&0&0&1&0\\
    0&0&0&-1&0&0&0&0&0&1\\
    0&0&1&0&0&0&0&0&-1&0\\
    0&0&0&0&0&-1&0&0&0&0\\
    0&0&0&0&1&0&0&0&1&0\\
    0&0&0&0&0&0&0&-1&0&0\\
    0&0&0&0&0&0&1&0&-1&0\\
    0&-1&0&1&0&-1&0&1&0&0\\
    1&0&-1&0&0&0&0&0&0&0

    \end{bmatrix}.
\end{equation}
This matrix is non-singular and its inverse is given by 
\begin{equation}
    (f_{ij}^{(2)})^{-1}=
    \begin{bmatrix}

    0 & \frac{1}{2} & 0 & \frac{1}{2} & 0& 0& 0&0&0&\frac{1}{2}\\
    -\frac{1}{2}&0&-\frac{1}{2}&0&\frac{1}{2}&0&-\frac{1}{2}&0&-\frac{1}{2}&0\\
    0&\frac{1}{2}&0&\frac{1}{2}&0&0&0&0&0&-\frac{1}{2}\\
    -\frac{1}{2}&0&-\frac{1}{2}&0&-\frac{1}{2}&0&\frac{1}{2}&0&\frac{1}{2}&0\\
    0&-\frac{1}{2}&0&\frac{1}{2}&0&1&0&0&0&\frac{1}{2}\\
    0&0&0&0&-1&0&0&0&0&0\\
    0&\frac{1}{2}&0&-\frac{1}{2}&0&0&0&1&0&-\frac{1}{2}\\
    0&0&0&0&0&0&-1&0&0&0\\
    0&\frac{1}{2}&0&-\frac{1}{2}&0&0&0&0&0&-\frac{1}{2}\\
    -\frac{1}{2}&0&\frac{1}{2}&0&-\frac{1}{2}&0&\frac{1}{2}&0&\frac{1}{2}&0
   
    \end{bmatrix}.
\end{equation}
The basic brackets of the theory can be obtained from the components of the inverse matrix as follows
\begin{eqnarray}
&&\{y_1,p_1\}=\frac{1}{2},\quad \{y_1,p_2\}=\frac{1}{2},\quad
\{y_1,\rho\}=\frac{1}{2},\quad \{y_3,p_1\}=-\frac{1}{2},\quad\{y_2,p_1\}=\frac{1}{2},\nonumber\\
&&\{y_2,p_2\}=\frac{1}{2},\quad
\{y_4,p_2\}=-\frac{1}{2},\quad \{p_2,\lambda\}=\frac{1}{2},\quad\{y_3,p_2\}=\frac{1}{2},\quad\{y_3,p_3\}=1,\nonumber\\
&&\{y_3,\rho\}=\frac{1}{2},\quad \{y_4,p_1\}=\frac{1}{2},\quad\{y_4,p_4\}=1,\quad\{\lambda,p_1\}=\frac{1}{2},\quad
\{y_2,\rho\}=-\frac{1}{2},\nonumber \\
&& \{y_4,\rho\}=-\frac{1}{2},\quad
\{\rho,\lambda\}=\frac{1}{2}.
\end{eqnarray} 
These brackets are consistent with the Dirac brackets of the theory$^*$.
\begingroup
\renewcommand\thefootnote{}
\footnotetext{$^*$ The brackets including $y_{1}$ and $p_{1}$ do not appear in \cite{brown2023singular} as these variables have been eliminated using constraint equation.}
\endgroup

The first-iterated symplectic matrix ($f_{ij}^{(1)}$) indicates that the given theory is a gauge theory. Therefore, the zero-modes $(\nu_k^{(1)})$ of this matrix will act as generators of gauge transformations in the following manner 
\begin{equation}\label{26}
    \delta\zeta_k^{(1)}=\nu_k^{(1)}\kappa,
\end{equation}
where $\kappa(t)$ denotes the time-dependent infinitesimal gauge parameter, and
\begin{equation}
    (\nu_{k}^{(1)})^{T}=(-1, 0, 1, 0, -1, 0, 1, 0, 1).
\end{equation}
Using relation \eqref{26}, we obtain the gauge transformations of the given system as follows:
\begin{equation}
\begin{aligned}
    &\delta y_1=-\kappa(t),  &\delta y_2=\kappa(t), & &\delta y_3=-\kappa(t),  &&\delta y_4=\kappa(t),  &&\delta\lambda=\kappa(t).
\end{aligned}
\end{equation}
The variation of $L^{(1)}$ under the above gauge transformations is given by
\begin{equation}
    \delta L^{(1)} = -\dot{\kappa}(t)[p_1-p_2+p_3-p_4],
\end{equation}
thus the Lagrangian $(L^{(1)})$ remains invariant on the constraint surface.

\subsection{Masses, Springs and a Ring}
\label{sub2}
We consider three identical masses sliding without friction on a ring of radius $R$. The masses are connected by springs. The system is defined by the Lagrangian
%\begin{figure}
%[h]
    %\centering
    %\begin{tikzpicture}[scale=3,>=stealth]
%[H]
%\draw[ultra thick, brown] (0,0) circle(1);
%\foreach \angle/\i in {40/1, 160/2, 290/3} {
    %\coordinate (m\i) at (\angle:1);
    %\fill[teal!70!black] (m\i) circle(0.15);
    %\node[above right=5pt] at (m\i) {$\theta_{\i}$};

    %\node[below right=8pt ] at (m\i) {$m_{\i}$};
%}
%\foreach \i in {1,2,3} {
    %\draw[decorate, decoration={coil, aspect=0.7, segment length=3mm, amplitude=1.9mm}, black, thick] (m\i) -- (0,0);
%}

%\draw[->] (0,0.3) -- (0,1.5) node[above] {\(y\)};
%\draw[->] (0.3,0) -- (1.5,0) node[right] {\(x\)};
%\draw[->] (20:1.3) arc (20:80:1.1);

%\end{tikzpicture}
   
    %\caption{\footnotesize{Three masses connected by springs, sliding without friction on a ring of radius $R$. }}
    %\label{fig2}
%\end{figure}
\begin{equation}
    \Tilde{L} = \frac{mR^2}{2} \left( \dot{\theta}_1^2 + \dot{\theta}_2^2 + \dot{\theta}_3^2 \right) - \Tilde{V}(\theta, x, y),
\end{equation}
where $\theta_1$, $\theta_2$ and $\theta_3$ are angles with the masses and $x$ and $y$ are the generalized coordinates of the system (cf.\cite{brown2023singular} for details).
The potential energy is given as
\begin{equation}
\begin{aligned}
    \Tilde{V}(\theta, x, y) &= \frac{k}{2} \bigg[
        (x - R \cos\theta_1)^2 + (y - R \sin \theta_1)^2 +
        (x - R \cos \theta_2)^2 \\&+ (y - R \sin \theta_2)^2 +
        (x - R \cos \theta_3)^2 + (y - R \sin \theta_3)^2
    \bigg].
\end{aligned}
\end{equation}
The conjugate momenta are obtained as follows
\begin{equation}
    \begin{aligned}
        &p_1 = mR^2 \dot{\theta}_1,\quad
        p_2 = mR^2 \dot{\theta}_2,\quad
        p_3 = mR^2 \dot{\theta}_3,\quad
        p_x = 0, \quad p_y = 0.
    \end{aligned}
\end{equation}
Here, $\Tilde{\phi_1}\equiv p_x=0$ and $\Tilde{\phi_2}\equiv p_y=0$ are the primary constraints in the context of Dirac's prescription for classification of constraints. The canonical Hamiltonian $(\Tilde{H}_c)$ can be obtained as 
\begin{equation}
    \Tilde{H}_c=\frac{1}{2mR^2}(p_1^2+p_2^2+p_3^2)+\Tilde{V}(\theta,x,y).
\end{equation}
The first-order Lagrangian takes the following form
\begin{equation}
    \Tilde{L}^{(0)}=p_1\dot{\theta_1}+p_2\dot{\theta_2}+p_3\dot{\theta_3}-\Tilde{V}^{(0)},
\end{equation}
with symplectic potential $\Tilde{V}^{(0)}=\frac{1}{2mR^2}(p_1^2+p_2^2+p_3^2)+\Tilde{V}(\theta,x,y)$.
The set of symplectic variables $(\Tilde{\zeta}^{(0)})$ of the theory is given as
\begin{equation}
    \Tilde{\zeta}^{(0)}=\{\theta_1,p_1,\theta_2,p_2,\theta_3,p_3,x,y\}.
\end{equation}
The calculation of symplectic one-forms results as
\begin{equation}
    \begin{aligned}
        &\Tilde{a}_1^{(0)}=p_1, \quad \Tilde{a}_2^{(0)}=0,
        \quad \Tilde{a}_3^{(0)}=p_2,\quad \Tilde{a}_4^{(0)}=0,\\
        &\Tilde{a}_5^{(0)}=p_3, \quad \Tilde{a}_6^{(0)}=0,
        \quad \Tilde{a}_7^{(0)}=0, \quad \Tilde{a}_8^{(0)}=0.
         \end{aligned}
\end{equation}
The corresponding symplectic matrix is constructed in the following fashion
\begin{equation}
    \Tilde{f}_{ij}^{(0)}=
    \begin{bmatrix}
        0&-1&0&0&0&0&0&0\\
        1&0&0&0&0&0&0&0\\
        0&0&0&-1&0&0&0&0\\
        0&0&1&0&0&0&0&0\\
        0&0&0&0&0&-1&0&0\\
        0&0&0&0&1&0&0&0\\
        0&0&0&0&0&0&0&0&\\
        0&0&0&0&0&0&0&0&
    \end{bmatrix}.
\end{equation}
This is a singular matrix, indicating the presence of constraints in the theory.
The zero-modes of the above matrix are found to be
\begin{equation}
\begin{aligned}
     &(\Tilde{\nu}_1^{(0)})^{T}=\{0,0,0,0,0,0,\Tilde{\nu}_1,0\}, \quad \text{and}\quad
     &(\Tilde{\nu}_2^{(0)})^{T}=\{0,0,0,0,0,0,0,\Tilde{\nu}_2\}.
\end{aligned}
\end{equation}
Contracting these zero-modes to the gradient of potential
\begin{equation}
    \frac{\partial {\Tilde{V}^{(0)}}(\Tilde{\zeta})}{\partial {\Tilde{\zeta}^{(0)}}}=
    \begin{bmatrix}
        kR(x\sin{\theta_1}-y\cos{\theta_1})\\
        \frac{p_1}{mR^2}\\
        kR(x\sin{\theta_2}-y\cos{\theta_2})\\
        \frac{p_2}{mR^2}\\
        kR(x\sin{\theta_3}-y\cos{\theta_3})\\
        \frac{p_3}{mR^2}\\
        k\{3x-R(\cos{\theta_1}+\cos{\theta_2}+\cos{\theta_3})\}\\
        k\{3y-R(\sin{\theta_1}+\sin{\theta_2}+\sin{\theta_3})\}
    \end{bmatrix},
\end{equation}
give rise to new constraints in the theory as follows
\begin{equation}
    \begin{aligned}
        &\Tilde{\Omega}_1^{(0)}=k\big[3x-R(\cos{\theta_1}+\cos{\theta_2}+\cos{\theta_3})\big],\\
        &\Tilde{\Omega}_2^{(0)}=k\big[3y-R(\sin{\theta_1}+\sin{\theta_2}+\sin{\theta_3})\big].
    \end{aligned}
\end{equation}
To check the existence of more constraints, we make use of the modified Faddeev-Jackiw formalism. Constructing the symplectic matrix $( \Tilde{f}_{kj}^{(1)})$ as in (\ref{eq:3}),
\begin{equation}
    \Tilde{f}_{kj}^{(1)}=
    \begin{bmatrix}
        0&-1&0&0&0&0&0&0\\
        1&0&0&0&0&0&0&0\\
        0&0&0&-1&0&0&0&0\\
        0&0&1&0&0&0&0&0\\
        0&0&0&0&0&-1&0&0\\
        0&0&0&0&1&0&0&0\\
        0&0&0&0&0&0&0&0&\\
        0&0&0&0&0&0&0&0&\\
        kR\sin{\theta_1}&0&kR\sin{\theta_2}&0&kR\sin{\theta_3}&0&3k&0\\
        -kR\cos{\theta_1}&0&-kR\cos{\theta_2}&0&-kR\cos{\theta_3}&0&0&3k
    \end{bmatrix}.
\end{equation}
The zero-mode of the matrix $(\Tilde{f}_{kj}^{(1)})$ can be obtained as $(\Tilde{\nu}^{(1)})^T=\{0,0,0,0,0,0,\Tilde{\nu}_1,\Tilde{\nu}_2,0,0\}$.\\
Multiplying this zero-mode with Eq.(\ref{eq:2}) and imposing the condition $\Tilde{\Omega}_1^{(0)}=\Tilde{\Omega}_2^{(0)}=0$ will indicate the existence of more constraints, if any. i.e.,
 \begin{equation}
    (\Tilde{\nu}^{(1)})^T \Tilde{Z}_k(\Tilde{\zeta}) \big|_{\Tilde{\Omega}_1^{(0)} = 0,\Tilde{\Omega}_2^{(0)}=0} = 0.
\end{equation} 
However, this relation gives identity, indicating the absence of any more constraints in the theory. The obtained constraints can now be introduced into the first-order Lagrangian using Lagrange multipliers $\lambda_1$ and $\lambda_2$.
Now, the Lagrangian takes the form
\begin{equation}
\begin{aligned}
    \Tilde{L}^{(1)}&=p_1\dot{\theta_1}+p_2\dot{\theta_2}+p_3\dot{\theta_3}+\dot{\lambda_1}k\{3x-R(\cos{\theta_1}+\cos{\theta_2}+\cos{\theta_3})\}\\
    &+\dot{\lambda_2}k\{3y-R(\sin{\theta_1}+\sin{\theta_2}+\sin{\theta_3})\}-\Tilde{V}^{(1)},
\end{aligned}
\end{equation}
where $\Tilde{V}^{(1)}=\Tilde{V}^{(0)} \big|_{\Tilde{\Omega}_1^{(0)} = 0,\Tilde{\Omega}_2^{(0)}=0}.$ The set of first-iterated variables can be identified as
\begin{equation}
    \Tilde{\zeta}^{(1)}=\{\theta_1,p_1,\theta_2,p_2,\theta_3,p_3,x,y,\lambda_1,\lambda_2\},
\end{equation}
and corresponding symplectic one-forms turn out be
\begin{align}
        &\Tilde{a}_1^{(1)}=p_1,\quad \Tilde{a}_2^{(1)}=0,\quad
        \Tilde{a}_3^{(1)}=p_2,\quad \Tilde{a}_4^{(1)}=0, \quad
        \Tilde{a}_5^{(1)}=p_3,\quad \Tilde{a}_6^{(1)}=0,
        \quad \Tilde{a}_x^{(1)}=0,\quad \Tilde{a}_y^{(1)}=0,\nonumber \\
        &\Tilde{a}_{\lambda_1}=k\{3x-R(\cos{\theta_1}+\cos{\theta_2}+\cos{\theta_3})\}, \quad
        \Tilde{a}_{\lambda_2}=k\{3y-R(\sin{\theta_1}+\sin{\theta_2}+\sin{\theta_3})\}.
\end{align}
Now, the first-iterated symplectic matrix can be constructed as
\begin{equation}
    \Tilde{f}_{ij}^{(1)}=
    \scalebox{0.9}{$
    \begin{bmatrix}
         0&-1&0&0&0&0&0&0&kR\sin{\theta_1}&-kR\cos{\theta_1}\\
        1&0&0&0&0&0&0&0&0&0\\
        0&0&0&-1&0&0&0&0&kR\sin{\theta_2}&-kR\cos{\theta_2}\\
        0&0&1&0&0&0&0&0&0&0\\
        0&0&0&0&0&-1&0&0&kR\sin{\theta_3}&-kR\cos{\theta_3}\\
        0&0&0&0&1&0&0&0&0&0\\
        0&0&0&0&0&0&0&0&3k&0\\
        0&0&0&0&0&0&0&0&0&3k\\
        -kR\sin{\theta_1}&0&-kR\sin{\theta_2}&0&-kR\sin{\theta_3}&0&-3k&0&0&0\\
        kR\cos{\theta_1}&0&kR\cos{\theta_2}&0&kR\cos{\theta_3}&0&0&-3k&0&0
    \end{bmatrix}$}.
\end{equation}
The matrix is non-singular and invertible. The inverse is given by
\begin{equation}
    (\Tilde{f}_{ij}^{(1)})^{-1}=
    \scalebox{0.9}{$
    \begin{bmatrix}
         0&1&0&0&0&0&0&0&0&0\\
        -1&0&0&0&0&0&\frac{R}{3}\sin{\theta_1}&-\frac{R}{3}\cos{\theta_1}&0&0\\
        0&0&0&1&0&0&0&0&0&0\\
        0&0&-1&0&0&0&\frac{R}{3}\sin{\theta_2}&-\frac{R}{3}\cos{\theta_2}&0&0\\
        0&0&0&0&0&1&0&0&0&0\\
        0&0&0&0&-1&0&\frac{R}{3}\sin{\theta_3}&-\frac{R}{3}\cos{\theta_3}&0&0\\
        0&-\frac{R}{3}\sin{\theta_1}&0&-\frac{R}{3}\sin{\theta_2}&0&-\frac{R}{3}\sin{\theta_3}&0&0&-\frac{1}{3k}&0\\
        0&-\frac{R}{3}\cos{\theta_1}&0&-\frac{R}{3}\cos{\theta_2}&0&-\frac{R}{3}\cos{\theta_3}&0&0&0&-\frac{1}{3k}\\
        0&0&0&0&0&0&\frac{1}{3k}&0&0&0\\
        0&0&0&0&&0&0&\frac{1}{3k}&0&0
    \end{bmatrix}$}.
\end{equation}
The components of the inverse matrix directly gives the basic brackets of the theory as follows
\begin{equation}
    \begin{aligned}
        &\{\theta_i,p_i\}=\delta_{ij},\quad \{x,p_i\}=-\frac{R}{3}\sin{\theta_i},\quad \{y,p_i\}=\frac{R}{3}\cos{\theta_i},\\
        &\{x,\lambda_1\}=-\frac{1}{3k},\quad \{y,\lambda_2\}=-\frac{1}{3k},
    \end{aligned}
\end{equation}
where $i=1,2,3.$ It can be observed that the basic brackets obtained via modified Faddeev-Jackiw quantization coincide with the Dirac brackets among the dynamical variables of the theory (cf.\cite{brown2023singular}).

\subsection{Pair of Pulleys}
\label{sub3}
The system in consideration comprises of three pairs of massless pulleys. Each upper pulley is connected to a mass and spring, while the lower pulleys are fixed. A continuous cord is threaded through the pulleys, looping over and under them. The left and right ends are identified such that the cord forms one continuous, closed loop (cf.\cite{brown2023singular} for details about the system).
We begin with the Lagrangian of the system, given by \cite{brown2023singular}
\begin{equation}\label{50}
\begin{aligned}
&L'=\frac{1}{8}mR^2[(\dot{\alpha_1} - \dot{\alpha_2})^2 + (\dot{\alpha_2} - \dot{\alpha_3})^2 + (\dot{\alpha_3} - \dot{\alpha_1})^2]-V'(\alpha),\\
\text{where}\qquad
&V'(\alpha) =\frac{1}{8}kR^2[(\alpha_1 - \alpha_2)^2 + (\alpha_2 - \alpha_3)^2 + (\alpha_3 - \alpha_1)^2].
\end{aligned}
\end{equation}
Here the angles $\alpha_{1}$,  $\alpha_{2}$  and  $\alpha_{3}$ of the lower fixed pulleys are the generalized coordinates of the system. As the cord moves over the pulleys, these angles change, affecting the heights of the masses. Each lower pulley has a radius $R$, whereas $m$ and $k$ denotes the mass and spring constant, respectively \cite{brown2023singular}. The canonical momenta are 
\begin{equation}
    \begin{aligned}
        &p_1=\frac{mR^2}{4}(2\dot{\alpha_1} - \dot{\alpha_2} - \dot{\alpha_3}),\quad 
        p_2=\frac{mR^2}{4}(2\dot{\alpha_2} - \dot{\alpha_3} - \dot{\alpha_1}),\\
        &p_3=\frac{mR^2}{4}(2\dot{\alpha_3} - \dot{\alpha_1} - \dot{\alpha_2}),
        \end{aligned}
\end{equation}
which leads to a single constraint in the theory
\begin{equation}
    \phi^{'} \equiv p_1+p_2+p_3=0. 
\end{equation}
The canonical Hamiltonian $(H'_c)$ of the system turns out to be
\begin{equation}
    H'_c = \frac{4}{3mR^2}(p_1^{2} + p_1p_2 + p_2^2) + V'(\alpha).
\end{equation}
As the Faddeev-Jackiw formalism requires the Lagrangian to be in the first-order form, therefore we convert Eq.\eqref{50} into the first-order one as
\begin{equation}
    L^{'(0)} = p_1\Dot{\alpha_1}+p_2\dot{\alpha_2}+p_3\dot{\alpha_3}-V^{'(0)},
\end{equation}
where 
\begin{equation}
V'^{(0)} = \frac{4}{3mR^2}(p_1^{2} + p_1p_2 + p_2^2) + V'(\alpha).    
\end{equation}
The constraint $\phi^{'}$ is introduced into the Lagrangian using a Lagrange multiplier $\lambda$ as follows:
\begin{equation}
L'^{(1)} = p_1\Dot{\alpha_1}+p_2\dot{\alpha_2}+p_3\dot{\alpha_3}+\dot{\lambda}(p_1+p_2+p_3)-V'^{(1)},
\end{equation}
where $V'^{(1)}=V'^{(0)}\vert_{\phi^{'}=0}$. Identifying the set of first-iterated symplectic variables of the theory as
\begin{equation}
    \zeta'^{(1)}=\{\alpha_1, p_1, \alpha_2, p_2, \alpha_3, p_3, \lambda\}.
\end{equation}
The components of the symplectic one-form are obtained as follows
\begin{equation}
\begin{aligned}
&a_{{1}}^{'(1)}=p_1, \quad a_{2}^{'(1)}=0, \quad a_{3}^{'(1)}=p_2, \quad a_{4}^{'(1)}=0,\\ 
&a_{5}^{'(1)}=p_3, \quad a_{6}^{'(1)}=0, \quad 
a_7^{'(1)}=p_1+p_2+p_3.
\end{aligned}
\end{equation}
The corresponding symplectic matrix takes the following form
\begin{equation}
f_{ij}^{'(1)}=
\begin{bmatrix}
    0 & -1 & 0 & 0 & 0& 0& 0\\
    1&0&0&0&0&0&1\\
    0&0&0&-1&0&0&0\\
    0&0&1&0&0&0&1\\
    0&0&0&0&0&-1&0\\
    0&0&0&0&1&0&1\\
    
    0&-1&0&-1&0&-1&0&
\end{bmatrix}.
\end{equation}
The singular nature of this matrix implies the presence of constraints in the theory. The zero-mode $(\nu'^{(1)})^T$ of the matrix turns out to be
\begin{equation}
    (\nu'^{(1)})^T=(-1, 0, -1, 0, -1, 0, 1).
    \label{eq.nu}
\end{equation}
To obtain new constraints, we contract this zero-mode with the equations of motion given in Eq.(\ref{eq:1}). This relation gives identity, indicating the absence of additional constraints. Since no further constraints arise and the matrix remains singular, the system exhibits a gauge symmetry.
The gauge condition we choose to quantize the theory is $\chi \equiv\alpha_1 + \alpha_2 + \alpha_3 =0$. This gauge condition is being incorporated into the Lagrangian using the Lagrange multiplier ($\rho$) as follows
\begin{equation}
    L'^{(2)}=p_1\dot{\alpha_1}+p_2\dot{\alpha_2}+p_3\dot{\alpha_3}+\dot{\lambda}(p_1+p_2+p_3)+\dot{\rho}(\alpha_1 + \alpha_2 + \alpha_3)-V'^{(2)},
\end{equation}
where $V'^{(2)}=V'^{(1)}\vert_{\chi=0}$. The set of second-iterated symplectic variables are 
\begin{equation}
    \zeta'^{(2)}=\{\alpha_1,p_1,\alpha_2,p_2,\alpha_3,p_3,\lambda,\rho\}.
\end{equation}
The respective symplectic one-forms turn out to be 
\begin{equation}
\begin{aligned}
&a_{1}^{'(2)}=p_1, \quad a_2^{'(2)}=0, \quad a_{3}^{'(2)}=p_2, \quad a_4^{'(2)}=0,\quad a_{5}^{'(2)}=p_3, \quad a_{6}^{'(2)}=0,\\ &a_{7}^{'(2)}=p_1+ p_2 + p_3, \quad a_{8}^{'(2)}=\alpha_1 + \alpha_2 + \alpha_3.
\end{aligned}
\end{equation}
The second-iterated symplectic matrix can be calculated as follows
\begin{equation}
f_{ij}^{'(2)}=
    \begin{bmatrix}
    0 & -1 & 0 & 0 & 0& 0& 0&1\\
    1&0&0&0&0&0&1&0\\
    0&0&0&-1&0&0&0&1\\
    0&0&1&0&0&0&1&0\\
    0&0&0&0&0&-1&0&1\\
    0&0&0&0&1&0&1&0\\
    
    0&-1&0&-1&0&-1&0&0\\
    -1&0&-1&0&-1&0&0&0

    \end{bmatrix}.
\end{equation}
This matrix is non-singular and its inverse is given by 
\begin{equation}
    (f^{'(2)}_{ij})^{-1}=
    \begin{bmatrix}

    0 & \frac{2}{3} & 0 & -\frac{1}{3} & 0& -\frac{1}{3}& 0&-\frac{1}{3}\\
    -\frac{2}{3}&0&\frac{1}{3}&0&\frac{1}{3}&0&-\frac{1}{3}&0\\
    0&-\frac{1}{3}&0&\frac{2}{3}&0&-\frac{1}{3}&0&-\frac{1}{3}\\
    \frac{1}{3}&0&-\frac{2}{3}&0&\frac{1}{3}&0&-\frac{1}{3}&0\\
    0&-\frac{1}{3}&0&-\frac{1}{3}&0&\frac{2}{3}&0&-\frac{1}{3}\\
    
    \frac{1}{3}&0&\frac{1}{3}&0&-\frac{2}{3}&0&-\frac{1}{3}&0\\
    
    0&\frac{1}{3}&0&\frac{1}{3}&0&\frac{1}{3}&0&\frac{1}{3}\\
    \frac{1}{3}&0&\frac{1}{3}&0&\frac{1}{3}&0&-\frac{1}{3}&0
   
    \end{bmatrix}.
\end{equation}
The basic brackets of the theory can be obtained from the components of the inverse matrix as follows
\begin{equation}
\begin{aligned}
&\{\alpha_i,p_j\}=\frac{2}{3} \quad;\quad i=j, \quad
\{\alpha_i,p_j\}=-\frac{1}{3} \quad; \quad i \neq j,\\
&\{\alpha_i,\rho\}=-\frac{1}{3},  \quad \{p_i,\lambda\} = -\frac{1}{3},\quad
\{\lambda,\rho\}=\frac{1}{3},
\end{aligned}    
\end{equation}\\
where $i, j =1,2,3.$ These brackets are consistent with the Dirac brackets of the theory \cite{brown2023singular}. The zero-mode $(\nu^{'(1)})^T$ of the singular symplectic matrix $(f_{ij}^{'(1)})$ act as the generators of gauge transformations.
Thus, the gauge transformations of the given system are obtained as follows:
\begin{equation}
\begin{aligned}
    &\delta' \alpha_1=-\kappa(t),  &\delta' \alpha_2=-\kappa(t), & &\delta' \alpha_3=-\kappa(t),&  &\delta'[p_1,p_2,p_3] = 0, &&\delta'\lambda=\kappa(t).
\end{aligned}
\end{equation}
The variation of the first-order Lagrangian is 
\begin{equation}
   \delta'{L'}^{(0)} = -\dot{\kappa}(t)[p_1+p_2+p_3], 
\end{equation}
i.e., the first-order Lagrangian $({L'}^{(0)})$ remains invariant on the constraint surface under this set of gauge transformations.

\section{Conclusions}
\label{con}

We have accomplished the quantization of three classical constrained systems \textit{\'{a} la} Faddeev-Jackiw formalism. The system of masses, rods and springs, and the coupled mass-pulley system are endowed with a single constraint. We incorporated the obtained constraint into the Lagrangian which resulted in a two-form singular symplectic matrix. As there are no further constraints and the symplectic matrix being singular indicated the presence of a gauge theory. Therefore, to quantize the system, we have chosen an appropriate gauge condition. This, in turn,  resulted in the two-form symplectic matrix become non-singular. Further, we have obtained the basic brackets of the theory from the components of the inverse of this second-iterated symplectic matrix. Whereas, the form of gauge transformations have been deduced from the zero-mode of the first-iterated symplectic matrix. 

For the system consisting of masses, springs
and a ring, we employed the modified Faddeev-Jackiw method to obtain new constraints. These constraints were incorporated into the Lagrangian, and the resulting matrix turned out to be non-singular. Further, we obtained the basic brackets from the inverse of the first-iterated symplectic matrix.

We observe that in each cases, the basic brackets obtained by means of Faddeev-Jackiw formalism are consistent with the Dirac brackets. However, in the Faddeev-Jackiw formalism, unlike the Dirac method, there are additional brackets that involve Lagrange multipliers. Thus, {\it{new}} interpretations were provided for the Lagrange multipliers in terms of the ``physical'' coordinates of the system. 

These results highlight the reliability of the Faddeev-Jackiw formalism as an alternative approach for the quantization of constrained systems.

\section*{Appendix A: On Lagrange Multipliers}
\subsection*{Masses, Rods and Springs}
We have incorporated constraints into the theory with the help of Lagrange multipliers. We have also obtained basic brackets including them. These brackets, involving Lagrange multipliers, do not appear in the Dirac formalism. Thus, to provide a {\it new} interpretation for Lagrange multipliers, we solve the symplectic equations of motion and find that they can be expressed in terms of physical coordinates of the theory. Thus, using \eqref{eq:1}, the equations of motion is given by
mm\begin{equation}
\scalebox{0.9}{$
\begin{bmatrix}
    \dot{y_1}\\
    \dot{p_1}\\
    \dot{y_2}\\
    \dot{p_2}\\
    \dot{y_3}\\
    \dot{p_3}\\
    \dot{y_4}\\
    \dot{p_4}\\
    \dot{\lambda}\\
    \dot{\rho}
\end{bmatrix}$}=
\scalebox{0.9}{$
    \begin{bmatrix}
    0 & \frac{1}{2} & 0 & \frac{1}{2} & 0& 0& 0&0&0&\frac{1}{2}\\
    -\frac{1}{2}&0&-\frac{1}{2}&0&\frac{1}{2}&0&-\frac{1}{2}&0&-\frac{1}{2}&0\\
    0&\frac{1}{2}&0&\frac{1}{2}&0&0&0&0&0&-\frac{1}{2}\\
    -\frac{1}{2}&0&-\frac{1}{2}&0&-\frac{1}{2}&0&\frac{1}{2}&0&\frac{1}{2}&0\\
    0&-\frac{1}{2}&0&\frac{1}{2}&0&1&0&0&0&\frac{1}{2}\\
    0&0&0&0&-1&0&0&0&0&0\\
    0&\frac{1}{2}&0&-\frac{1}{2}&0&0&0&1&0&-\frac{1}{2}\\
    0&0&0&0&0&0&-1&0&0&0\\
    0&\frac{1}{2}&0&-\frac{1}{2}&0&0&0&0&0&-\frac{1}{2}\\
    -\frac{1}{2}&0&\frac{1}{2}&0&-\frac{1}{2}&0&\frac{1}{2}&0&\frac{1}{2}&0
   
    \end{bmatrix}$}
    \scalebox{0.9}{$
    \begin{bmatrix}
    -mg + \frac{k}{4}[4a-2y_1-y_2-y_4]\\
    \frac{1}{2m}[5p_1-2p_2-2p_4+p_3]\\
    -mg+\frac{k}{4}[4a-y_1-2y_2-y_3]\\
    \frac{1}{2m}[5p_2-2p_1-2p_3+p_4]\\
    -mg+\frac{k}{4}[4a-y_2-2y_3-y_4]\\
    \frac{1}{2m}[5p_3-2p_2-2p_4+p_1]\\
    -mg+\frac{k}{4}[4a-y_3-2y_4-y_1]\\
    \frac{1}{2m}[5p_4-2p_3-2p_1+p_2]\\
    0\\
    0
    \end{bmatrix}$}.
\end{equation}
From the above relation we obtain, 
\begin{equation}
\begin{aligned}
    &\dot{\lambda}=\frac{1}{4m}[7p_1-7p_2+3p_3-3p_4],\quad
    &\dot{\rho}=0.
\end{aligned}
\end{equation}
We can see that one of the Lagrange multipliers can be represented in terms of the momenta of the theory.
\label{sub1}

\subsection*{Masses, Springs and a Ring}
In case of a system containing masses, springs and ring, solving the symplectic equations of motion, yeilds
\begin{equation}\label{71}
\scalebox{0.85}{$
\begin{bmatrix}
    \dot{\theta_1}\\
    \dot{p_1}\\
    \dot{\theta_2}\\
    \dot{p_2}\\
    \dot{\theta_3}\\
    \dot{p_3}\\
    \dot{x}\\
    \dot{y}\\
    \dot{\lambda_1}\\
    \dot{\lambda_2}
\end{bmatrix}$}=
\scalebox{0.85}{$
    \begin{bmatrix}
         0&1&0&0&0&0&0&0&0&0\\
        -1&0&0&0&0&0&\frac{R}{3}\sin{\theta_1}&-\frac{R}{3}\cos{\theta_1}&0&0\\
        0&0&0&1&0&0&0&0&0&0\\
        0&0&-1&0&0&0&\frac{R}{3}\sin{\theta_2}&-\frac{R}{3}\cos{\theta_2}&0&0\\
        0&0&0&0&0&1&0&0&0&0\\
        0&0&0&0&-1&0&\frac{R}{3}\sin{\theta_3}&-\frac{R}{3}\cos{\theta_3}&0&0\\
        0&-\frac{R}{3}\sin{\theta_1}&0&-\frac{R}{3}\sin{\theta_2}&0&-\frac{R}{3}\sin{\theta_3}&0&0&-\frac{1}{3k}&0\\
        0&-\frac{R}{3}\cos{\theta_1}&0&-\frac{R}{3}\cos{\theta_2}&0&-\frac{R}{3}\cos{\theta_3}&0&0&0&-\frac{1}{3k}\\
        0&0&0&0&0&0&\frac{1}{3k}&0&0&0\\
        0&0&0&0&0&0&0&\frac{1}{3k}&0&0
    \end{bmatrix}$}
    \frac{\partial\Tilde{V}^{(1)}(\Tilde{\zeta})}{\partial\Tilde{\zeta}^{(1)}},
\end{equation}
where 
\begin{equation}\label{72}
\frac{\partial\Tilde{V}^{(1)}(\Tilde{\zeta})}{\partial\Tilde{\zeta}^{(1)}}=
\begin{bmatrix}
   kR(x\sin{\theta_1}-y\cos{\theta_1})\\
        \frac{p_1}{mR^2}\\
        kR(x\sin{\theta_2}-y\cos{\theta_2})\\
        \frac{p_2}{mR^2}\\
        kR(x\sin{\theta_3}-y\cos{\theta_3})\\
        \frac{p_3}{mR^2}\\
        k\{3x-R(\cos{\theta_1}+\cos{\theta_2}+\cos{\theta_3})\}\\
        k\{3y-R(\sin{\theta_1}+\sin{\theta_2}+\sin{\theta_3})\}\\
        0\\0
\end{bmatrix}.
\end{equation}
Thus, from above equations \eqref{71} and \eqref{72}, we have,
\begin{eqnarray}
    \dot{\lambda_1}=x-\frac{R}{3}(\cos{\theta_1}+\cos{\theta_2}+\cos{\theta_3}), \qquad
    \dot{\lambda_2}=y-\frac{R}{3}(\sin{\theta_1}+\sin{\theta_2}+\sin{\theta_3}).
    \end{eqnarray}
i.e., the Lagrange multipliers can be represented in terms of the coordinates of the theory.

\subsection*{Pair of Pulleys}
Finally, in case of pair of pulleys, symplectic equations of motion can be written as
\[
\scalebox{0.9}{$
\begin{bmatrix}
    \dot{\alpha_1}\\
    \dot{p_1}\\
    \dot{\alpha_2}\\
    \dot{p_2}\\
    \dot{\alpha_3}\\
    \dot{p_3}\\
    
    \dot{\lambda}\\
    \dot{\rho}
\end{bmatrix}$}=
\scalebox{0.9}{$  
   \begin{bmatrix}
    0 & \frac{2}{3} & 0 & -\frac{1}{3} & 0& -\frac{1}{3}& 0&-\frac{1}{3}\\
    -\frac{2}{3}&0&\frac{1}{3}&0&\frac{1}{3}&0&-\frac{1}{3}&0\\
    0&-\frac{1}{3}&0&\frac{2}{3}&0&-\frac{1}{3}&0&-\frac{1}{3}\\
    \frac{1}{3}&0&-\frac{2}{3}&0&\frac{1}{3}&0&-\frac{1}{3}&0\\
    0&-\frac{1}{3}&0&-\frac{1}{3}&0&\frac{2}{3}&0&-\frac{1}{3}\\
    
    \frac{1}{3}&0&\frac{1}{3}&0&-\frac{2}{3}&0&-\frac{1}{3}&0\\
    
    0&\frac{1}{3}&0&\frac{1}{3}&0&\frac{1}{3}&0&\frac{1}{3}\\
    \frac{1}{3}&0&\frac{1}{3}&0&\frac{1}{3}&0&-\frac{1}{3}&0
   
    \end{bmatrix}$}
    \scalebox{0.9}{$
    \begin{bmatrix}
    -\frac{kR^2}{4}[2\alpha_1-\alpha_2-\alpha_3]\\
    \frac{4}{3mR^2}[2p_1+p_2]\\
    -\frac{kR^2}{4}[2\alpha_2-\alpha_3-\alpha_1]\\
    \frac{4}{3mR^2}[p_1+2p_2]\\
    -\frac{kR^2}{4}[2\alpha_3-\alpha_1-\alpha_2]\\
    0\\
    0\\
    0
    \end{bmatrix}$}.
    \]
The above relation yields the following relations: 
\begin{equation}
\begin{aligned}
    &\dot{\lambda}=\frac{4}{3mR^2}[p_1+p_2],\quad
    \dot{\rho}=0.
\end{aligned}
\end{equation}
It can be observed that one of the Lagrange multipliers is zero, while the other can be represented in terms of the momenta of the theory.

 \section*{Appendix B: MATLAB Implementation of Faddeev-Jackiw Quantization} 
The Faddeev-Jackiw quantization process for singular Lagrangian systems is computationally realized in this section. The MATLAB code is designed to manage the symplectic formulation of the systems under study, allowing for automated treatment of constraints, symplectic matrix construction, and basic bracket deduction. Below are the pseudocode and thorough descriptions of the structure and logic of the algorithm.

\begin{algorithm}[H]
\caption{System Initialization}
\begin{algorithmic}[1]
\Function{FaddeevJackiwQuantization}{$V$, $L$, $q$, $p$}
\State Prompt the user to input potential and Lagrangian
\State Define generalized coordinates
\For{$i \gets 1$ \textbf{to} $n$} \Comment{Momentum Calculation}
    \State $p_i \gets \frac{\partial L}{\partial \dot{q}_i}$
    \State Solve $\dot{q}_i$ from $p_i = \frac{\partial L}{\partial \dot{q}_i}$
\EndFor
\State $H \gets \sum\limits_{i=1}^n p_i\dot{q}_i - L$ \Comment{Hamiltonian Formulation}
\For{$i \gets 1$ \textbf{to} $\mathrm{length}(p)$}
    \If{$p_i \neq 0$} 
        \State $H \gets \left.H\right|_{\frac{\partial L}{\partial \dot{q}_i} \rightarrow p_i}$ 
    \EndIf
\EndFor
\State $L^{(0)} \gets \sum\limits_{i=1}^n p_i\dot{q}_i - H$ \Comment{First-order Lagrangian}
\If{$\forall\, i,\; p_i \neq 0$} \Comment{Check for inherent constraints} 
    \State $\Omega^{(0)} \gets \texttt{input('Inherent constraint: ')}$ 
    \State $L_1 \gets L_0 + \sum\limits_{i=1}^m \dot{\lambda}_i\Omega^{(0)}_i$ 
\Else
    \State $L_1 \gets L_0$ 
\EndIf
    
\EndFunction
\end{algorithmic}
\end{algorithm}
\begin{itemize}
 \item {\bf{User Input and System Initialization}}\\
    The user can select from a list of predefined variable sets, each
corresponding to a particular physical system. Only one such line must be active
at a time. The user is prompted to input the generalized coordinates, the potential, and the Lagrangian.\\
    For each coordinate, the canonical momentum is computed and the Hamiltonian and first-order Lagrangian is constructed. If the system does not exhibit any inherent (primary) constraints, the user must input a constraint relation among the canonical momenta which is incorporated into the Lagrangian before proceeding.
    \item {\bf{Symplectic Structure Calculation}}\\
    The function \texttt{calculateSymplecticStructure} determines the set of symplectic variables, calculates the symplectic one-form components, and builds the symplectic matrix. It then determines the determinant to look for singularity.
\begin{algorithm}[H]
\caption{Symplectic Structure Calculation}
\begin{algorithmic}[1]
\Function{calculateSymplecticStructure}{$L_1$, \textit{vars}, $p$, \textit{dq}}
    \State Find intersection: $\zeta \gets \textit{vars} \cap \text{symvar}(L_1)$ \Comment{Determine symplectic variables}
    \State Create differentials: $\textit{d}\zeta \gets \{d\zeta_1, d\zeta_2,...\}$ 
    \For{$i \gets 1$ to $\text{length}(d\zeta)$}
        \State $a_i \gets \frac{\partial L_1}{\partial (d\zeta_i)}$
        
    \EndFor
    
    \For{$i,j \gets 1$ to  $\text{length}(d\zeta)$} \Comment{Symplectic matrix calculation}
        \State $f_{ij} \gets \frac{\partial a_j}{\partial \zeta_i} - \frac{\partial a_i}{\partial \zeta_j}$
    \EndFor

    \State Find $\det(f)$ 
    
    \State \Return $\zeta, d\zeta, a, F, \det F$
\EndFunction
\end{algorithmic}
\end{algorithm}

    \item {\bf{Constraint Detection and Gauge Fixing}}\\
   If the matrix is singular, the code finds its zero-modes and uses the gradient of potential to contract them in order to look for new constraints. It updates the Lagrangian iteratively with these constraints until no further constraints are obtained. If the matrix remains singular after all the constraints are obtained, the user is prompted for gauge-fixing conditions, which are incorporated as new constraints, and the symplectic structure is recalculated.
\begin{algorithm}[H]
\caption{Constraint Handling and Gauge Fixing}
\begin{algorithmic}[1]
\Function{FaddeevJackiwQuantization}{$V$, $L$, $q$, $p$}
\While{$\det(f) = 0$}
        \State Find zero modes $v^{(\alpha)}$ of $f$
        \State Derive constraints $\Omega^{(\alpha)} = v^{(\alpha)} \cdot \nabla V^{(0)}$
        \State $L_1 = L_1 + \dot{\lambda}_\alpha\Omega^{(\alpha)}$
        \State Call function \textsc{calculateSymplecticStructure}  \Comment{Update symplectic structure}
\EndWhile
\If{$\det f = 0$} \Comment{Gauge fixing }
        \State $g \gets \texttt{input('Gauge condition: ')}$
        \State $L_2 = L_1 + \dot{\rho}g$
        \State Call function \textsc{calculateSymplecticStructure} 
    
    \EndIf
    \State $f^{-1} \gets \texttt{inv}(f)$
\EndFunction
\end{algorithmic}
\end{algorithm}
    \item {\bf{Quantization Brackets}}\\
    All non-zero brackets from the inverse of the non-singular symplectic matrix are extracted and displayed by the code.
\begin{algorithm}[H]
\caption{Quantization brackets}
\begin{algorithmic}[1]
\Function{FaddeevJackiwQuantization}{$V$, $L$, $q$, $p$}
\State $[m, n] \gets \text{size}(F^{-1})$
\For{$i \gets 1$ \textbf{to} $m$}
    \For{$j \gets i+1$ \textbf{to} $n$}
        \State $f_{ij} \gets F^{-1}(i,j)$
        \If{$f_{ij} \neq 0$}
            \State \text{Output} ``\{$\zeta_i$, $\zeta_j$\} = $f_{ij}$ = -\{$\zeta_j$, $\zeta_i$\}''
        \EndIf
    \EndFor
\EndFor
\EndFunction
\end{algorithmic}
\end{algorithm}
\end{itemize}
The code handles user-defined Lagrangians dynamically. Only system-specific information needs user input because constraints are automatically extracted and incorporated. User-supplied conditions are used to identify and resolve gauge freedom. At every stage, intermediate results are also shown.

%\newpage
\bibliographystyle{unsrt}
\bibliography{biblio}

@article{Anjali:2019ocq,
    author = "S. Anjali and S. Gupta",
    title = "{Faddeev\textendash{}Jackiw quantization of Christ\textendash{}Lee model}",
    eprint = "1908.05499",
    archivePrefix = "arXiv",
    primaryClass = "hep-th",
    doi = "10.1142/S0217732320500728",
    journal = "Mod. Phys. Lett. A",
    volume = "35",
    
    pages = "2050072",
    year = "2020"
}

@book{dirac2001lectures,
  title="{Lectures on Quantum Mechanics}",
  author="{P. A. M. Dirac}",
  year={1964},
  publisher={Belfer Graduate School of Science, Yeshiva University}
}

@article{brown2023singular,
  title="{Singular Lagrangians and the Dirac-Bergmann algorithm in classical mechanics}",
  author="{J. D. Brown}",
  journal="{Am. J. Phys}",
  volume={91},
  
  pages={214--224},
  year={2023},
  publisher={AIP Publishing}
}

@article{HUANG2008438,
title = "{Modified Faddeev–Jackiw quantization of massive non-Abelian Yang–Mills fields and Lagrange multiplier fields}",
journal = "{Phys. Lett. B}",
volume = {668},

pages = {438-441},
year = {2008},
issn = {0370-2693},
doi = {https://doi.org/10.1016/j.physletb.2008.05.073},
url = {https://www.sciencedirect.com/science/article/pii/S0370269308011039},
author = "{Y. C. Huang and J. L. Yang}",

}

@article{doi:10.1142/S0217732323501869,
author = "{A. S. Nair and S. Gupta}",
title = "{On the quantization of FLPR model}",
journal = {Mod. Phys. Lett. A},
volume = {39},

pages = {2350186},
year = {2024},
doi = {10.1142/S0217732323501869},

URL = { 
    
        https://doi.org/10.1142/S0217732323501869
},
eprint = { 
    
        https://doi.org/10.1142/S0217732323501869
}
,
   
}

@article{Paulin-Fuentes:2023zya,
    author = "J. M. Paulin-Fuentes, C. M. L. Arellano and J. M. Cabrera",
    title = "{Singular Lagrangians and the Faddeev-Jackiw formalism in classical mechanics}",
    eprint = "2311.09407",
    archivePrefix = "arXiv",
    primaryClass = "math-ph",
    doi = "10.1007/s10773-024-05626-7",
    journal = "Int. J. Theor. Phys.",
    volume = "63",
   
    pages = "111",
    year = "2024"
}

@article{PhysRevLett.60.1692,
  title = "{Hamiltonian reduction of unconstrained and constrained systems}",
  author = {Faddeev, L. and Jackiw, R.},
  journal = {Phys. Rev. Lett.},
  volume = {60},
  issue = {17},
  pages = {1692},
  numpages = {0},
  year = {1988},
  
  publisher = {American Physical Society},
  doi = {10.1103/PhysRevLett.60.1692},
  url = {https://link.aps.org/doi/10.1103/PhysRevLett.60.1692}
}

@Article{universe8030171,
AUTHOR ="{J. D. Brown}",
TITLE = "{Singular Lagrangians, constrained Hamiltonian systems and gauge invariance: An example of the Dirac–Bergmann algorithm}",
JOURNAL = {Universe},
VOLUME = {8},
YEAR = {2022},

ARTICLE-NUMBER = {171},
URL = {https://www.mdpi.com/2218-1997/8/3/171},
ISSN = {2218-1997},

DOI = {10.3390/universe8030171}
}

@article{Foussats1997,
  author = {A. Foussats and C. Repetto and O. P. Zandron and O. S. Zandron},
  title = "{Nonlinear sigma model in the Faddeev-Jackiw quantization formalism}",
  journal = "{Int. J. Theor. Phys}",
  year = {1997},
  volume = {36},
 
  pages = {2923--2935},
  doi = {10.1007/BF02435718},
  url = {https://doi.org/10.1007/BF02435718}
}

@article{doi:10.1142/S0217732393003810,
author = "{H. Montani and C. Wotzasek}",
title = "{Faddeev-Jackiw quantization of non-Abelian systems}",
journal = {Mod. Phys. Lett. A},
volume = {08},

pages = {3387-3396},
year = {1993},
doi = {10.1142/S0217732393003810},

URL = { 
    
        https://doi.org/10.1142/S0217732393003810
},
eprint = { 
    
        https://doi.org/10.1142/S0217732393003810
}  
}

@article{RomeroHernandez2025,
  author    = "{L. G. R. Hernández, J. M. Cabrera, R. E. C. López and J. M. Paulin-Fuentes}",
  title     = "{Singular Lagrangians and the Hamilton-Jacobi formalism in classical mechanics}",
  journal   = "{Int. J. Theor. Phys.}",
  volume    = {64},
  
  pages     = {36},
  year      = {2025},
 
  doi       = {10.1007/s10773-025-05887-w},
  url       = {https://doi.org/10.1007/s10773-025-05887-w},
  issn      = {1572-9575},
  
}

@article{HUANG20102140,
title = "{Faddeev–Jackiw and the improved methods in quantization of the superconductive system}",
journal = "{Ann. Phys}",
volume = {325},

pages = {2140-2152},
year = {2010},
issn = {0003-4916},
doi = {https://doi.org/10.1016/j.aop.2010.05.010},
url = {https://www.sciencedirect.com/science/article/pii/S0003491610001016},
author = "{Y. C. Huang and L. X. Yi}",
}

@article{ESCALANTE2016375,
title = "{Faddeev–Jackiw quantization of four dimensional BF theory}",
journal = "{Ann. Phys}",
volume = {374},
pages = {375-394},
year = {2016},
issn = {0003-4916},
doi = {https://doi.org/10.1016/j.aop.2016.09.003},
url = {https://www.sciencedirect.com/science/article/pii/S0003491616301774},
author ="{A. Escalante and P. C. Sánchez}",

}

@article{Toms:2015lza,
    author = "D. J. Toms",
    title = "{Faddeev-Jackiw quantization and the path integral}",
    eprint = "1508.07432",
    archivePrefix = "arXiv",
    primaryClass = "hep-th",
    doi = "10.1103/PhysRevD.92.105026",
    journal = "Phys. Rev. D",
    volume = "92",
    
    pages = "105026",
    year = "2015"
}

@article{S:2021msx,
    author = "S. Anjali and S. Gupta",
    title = "{Particle on a torus knot: symplectic analysis}",
    eprint = "2108.08788",
    archivePrefix = "arXiv",
    primaryClass = "hep-th",
    doi = "10.1140/epjp/s13360-022-02699-3",
    journal = "Eur. Phys. J. Plus",
    volume = "137",
    number = "4",
    pages = "511",
    year = "2022"
}

@article{S:2021zob,
    author = "S, Anjali and Gupta, S.",
    title = "{Symplectic gauge-invariant reformulation of a free-particle system on toric geometry}",
    eprint = "2105.07742",
    archivePrefix = "arXiv",
    primaryClass = "hep-th",
    doi = "10.1209/0295-5075/135/11002",
    journal = "EPL",
    volume = "135",
    number = "1",
    pages = "11002",
    year = "2021"
}

@article{Gupta:2016qwv,
    author = "Gupta, S. and Roychowdhury, R.",
    title = "{Antiself-dual Yang{\textendash}Mills, modified Faddeev{\textendash}Jackiw formalism and hidden BRS invariance}",
    eprint = "1603.09299",
    archivePrefix = "arXiv",
    primaryClass = "hep-th",
    doi = "10.1142/S0217751X16501384",
    journal = "Int. J. Mod. Phys. A",
    volume = "31",
    number = "24",
    pages = "1650138",
    year = "2016"
}

@article{Barcelos-Neto:1991dhe,
    author = "Barcelos-Neto, J. and Wotzasek, C.",
    title = "{Symplectic quantization of constrained systems}",
    reportNumber = "IF-UFRJ-15-91",
    doi = "10.1142/S0217732392001439",
    journal = "Mod. Phys. Lett. A",
    volume = "7",
    pages = "1737--1748",
    year = "1992"
}

\end{document}